\documentclass[nocompress]{spie}  

\usepackage{amsmath,amsfonts,amssymb}
\def\arcmin{\hbox{$^\prime$}}
\def\arcsec{\hbox{$^{\prime\prime}$}}
\def\degrees{\ifmmode^\circ\else$^\circ$\fi}
\usepackage{graphicx}
\usepackage{float}
\usepackage{wrapfig}

\title{Enhanced Seeing Mode at the LBT: A Method to Significantly Improve Angular Resolution over a 4$\arcmin$ $\times$ 4$\arcmin$ Field of View}

\author[a,b]{Barry Rothberg}
\author[a]{Julian C. Christou}
\author[a]{Douglas L. Miller}
\author[a]{Dave Thompson}
\author[a]{Gregory E. Taylor}
\author[a]{Christian Veillet}
\affil[a]{Large Binocular Telescope Observatory, 933 North Cherry Avenue, Tucson, AZ 85721, USA}
\affil[b]{Department of Physics and Astronomy, George Mason University, MS 3F3, 4400 University Drive, Fairfax, VA 22030, USA}

\authorinfo{Further author information: (Send correspondence to Barry Rothberg)\\
E-mail: brothberg@lbto.org, Telephone: +1 520 626-8672, Fax: +1 520 626-9333}

\begin{document} 

\maketitle 

\begin{abstract}
Since 2014, the LBT's First Light Adaptive Optics (FLAO) system has also included a 
seldom used capability, known as Enhanced Seeing Mode (ESM), that can 
improve the angular resolution over a 4$\arcmin$ $\times$ 4$\arcmin$ field of view (FOV). In 
full AO operation, FLAO provides diffraction limited (DL) capabilities over a small 
(30\arcsec $\times$ 30\arcsec) FOV. By comparison, ESM can achieve significantly enhanced 
resolution, over natural seeing, across a far larger FOV. This improves operational 
efficiency over standard seeing limited (SL) observations and is applicable across a 
broader range of scientific targets. ESM uses 11 modes of correction (including tip and 
tilt) to remove residual aberrations and jitter which significantly improves angular 
resolution over the full FOV. While this mode does not reach the DL, it can achieve 
uniform angular resolutions as good as 0\arcsec.22 over the FOV. Furthermore, it allows 
for the use of multi-object spectroscopy with R$\sim$10,000 or imaging with angular 
resolution similar to that achieved by the Wide-Field Camera 3 infrared channel on the 
Hubble Space Telescope, but powered by 11.6 meters of effective aperture in binocular 
mode. As part of the on-going characterization of ESM, we have demonstrated that even in 
poor seeing conditions ($\theta$ $\sim$ 1\arcsec.5-2\arcsec) the image quality delivered to the 
focal station is improved by factors of 2-3. Here, we present the first results of the 
characterization of ESM, including systematic tests of the delivered PSF across the FOV as 
a function of the brightness of, and distance from, the AO Reference Star. We present a 
range of galactic and extra-galactic targets showing the improvements obtained over a broad 
range of seeing conditions and propose ESM as a standard observational mode for 
near-Infrared observations.
\end{abstract}


\section{INTRODUCTION}
\label{sec:intro}  
\subsection{The Large Binocular Telescope}\label{subsec:LBTO}
\indent The Large Binocular Telescope (LBT) is located on Emerald Peak (elevation 3,192 
meters) on Mount Graham, which is located in southeastern Arizona near the city of Safford. 
The observatory operates from September 1-July 10 each year.  Between July 11 and August 
31 the observatory closes for monsoon season and this down-time is used for telescope and 
instrument maintenance and upgrades.  The LBT is an international partnership which 
includes: the University of Arizona, Arizona State University and Northern Arizona 
University (25$\%$ share of time); Germany or LBT Beteiligungsgesellschaft (25$\%$ share 
of time), which includes the German institutes of Landessternwarte K{\"{o}nigstuhl, 
Leibniz Institute for Astrophysics Potsdam (AIP), Max-Planck-Institut f{\"{u}r Astronomie, 
Max-Planck-Institut f{\"{u}r Extraterrestrische Physik, and Max-Planck-Institut f{\"{u}r 
Radioastronomie. Some of this time is also allocated to public universities within Germany; 
Italy or Instituto Nazionale di Astrofisica (25$\%$ share of time), which offers access to 
the entire Italian astronomical community; The Ohio State University (12.5$\%$ share of 
time); and the Research Corporation for Science and Advancement (12.5$\%$ share of time) 
which coordinates the participation of four universities (The Ohio State University, 
University of Notre Dame, University of Minnesota, and University of Virginia).\\
\indent The Large Binocular Telescope houses two 8.4 meter primary mirrors, separated by 
14.4 meters (center-to-center), affixed to a single compact altitude-azimuth mount housed 
in a co-rotating enclosure, see Hill et al. (2004)\cite{2004SPIE.5489..603H}, 
Hill et al. (2010)\cite{2010SPIE.7733E..0CH} and references therein for more details.  
Each mirror has four Bent Gregorian focal stations, a direct Gregorian focal station,
and a swing arm which houses a prime focus optical camera. Another swing arm houses the 
adaptive secondary mirror which is placed into the beam when the prime focus camera is not 
being used.  A third swing arm moves and aligns tertiary mirror with the requested Bent 
Gregorian focal station. The transition between prime focus and Gregorian instruments 
takes $\sim$ 20 minutes, while transitions between different Gregorian instruments can 
take $\le$ 10 minutes.  The left-side of the telescope is denoted as {\it SX} (the `S' from 
the Italian word for left, {\it sinistra}), and the right-side is denoted as {\it DX} (the
'D' from the Italian word for right, {\it destra}).\\
\indent The unique design of LBT allows for its use in three configurations: 1) ``Twinned'' 
or Duplex mode in which each mirror has identical instrument configurations yielding an 
effective collecting area of $\simeq$ 11.6 meters; 2) Interferometric mode which uses
the baseline between mirrors to create an effective aperture of 22.6 meters; and 
3) ``Heterogeneous'' mode, in which the 8.4 meter primary mirrors are each configured with
different instruments or different modes of the same instruments (i.e. optical on SX, 
near-IR on DX, or optical imaging on SX and optical spectroscopy on DX) and operate as 
two independent telescopes on a common mount.  In this mode, the two mirrors
can move independently of each other up to the ``co-pointing limit'' ($\sim$ 40\arcsec),
(e.g. Rakich et al. 2011\cite{2011SPIE.8128E..08R}, 
Hill et al. 2014\cite{2014SPIE.9145E..02H}, 
Rothberg et al. 2018\cite{2018SPIE10702E..05R}).

\subsection{First Light Adaptive Optics (FLAO) System}\label{subsec:FLAO}
\indent The Natural Guide Star (NGS) adaptive optics system at LBT uses an Adaptive 
Secondary Mirror with 672 actuators driving a deformable shell.  A pyramid wavefront 
sensor uses the brightness and point-spread function (PSF) of an NGS to apply corrections 
for atmospheric turbulence. Currently, the First Light Adaptive Optics (FLAO) system 
offers from 36 to 400 modes of corrections for full diffraction limited (DL) AO. It can 
correct for non-common path aberrations (NCPA) using on-axis bright natural guide-stars 
(Esposito et al. 2010\cite{2010SPIE.7736E..09E}, 2012\cite{2012SPIE.8447E..0UE},
Christou et al. 2016 \cite{2016SPIE.9909E..2EC}, 2018\cite{2018SPIE10703E..0AC}, and
Miller et al. 2016\cite{2016SPIE.9909E..2GM}.  The corrections need to achieve a DL PSF
are directly related to the brightness of the NGS, the angular separation between
it and the science target, and the wavelength of the observations.\\
\indent However, FLAO can also apply 11 modes to improve angular 
resolution: tip, tilt, focus, 2 $\times$ astigmatism, 2 $\times$ coma, 2 $\times$ trefoil, 
spherical and 1 “high” order.  These 11 modes define Enhanced Seeing Mode (ESM). The AO 
patrol field used for both diffraction limited AO and ESM is 3$\arcmin$ $\times$ 2$\arcmin$ 
and requires a reference star {\it m}$_{\rm R}$ $\le$ 16.  Currently, the FLAO system is 
undergoing an upgrade to the next generation Single conjugated adaptive Optics Upgrade for 
LBT (SOUL).  This upgrade will increase the number of modes which can be applied to full 
DL AO and allow for use of AO Reference Stars 1.5-2 magnitudes fainter than the current 
limit (Pinna et al. 2016\cite{2016SPIE.9909E..3VP}, 
Christou et al. 2018\cite{2018SPIE10703E..0AC}).  The SOUL upgrade is 
undergoing commissioning on SX and the hardware will be installed on DX during the Fall 
and Winter of 2019.

\subsection{LBT Utility Camera in the Infrared (LUCI)}\label{subsec:LUCI}
\indent The two LBT Utility Camera in the Infrared instruments (LUCIs), are a pair of 
cryogenic near-Infrared (NIR) instruments sensitive from 
0.95\footnote{The dichroic dewar entrance window cut-on is $\sim$ 0.89$\mu$m for LUCI-1 on 
SX and $\sim$ 0.95$\mu$m for LUCI-2 on DX}-2.4$\mu$m with imaging and spectroscopic 
(longslit and multi object slit) capabilities.  Each LUCI houses a 2K $\times$ 2K Hawaii 
2RG detector and three cameras which deliver a 0\arcsec.25/pixel 
({\it f}/1.8 designated N1.8), 0\arcsec.12/ pixel ({\it f}/3.75 designated N3.75), or a 
0\arcsec.015/pixel ({\it f}/30 designated N30) plate scale. The N1.8 camera is only used 
for seeing limited (SL) spectroscopy.  The N30 camera is designed for DL AO and delivers a
30$\arcsec$ $\times$ 30$\arcsec$ FOV (currently only available in imaging mode).  The N3.75 
camera yields a 4$\arcmin$ $\times$ 4$\arcmin$ FOV and is used for both SL and ESM imaging 
and spectroscopy (DL AO is under-sampled with this camera).    Each LUCI is mounted at the 
front bent Gregorian port of SX or DX.  Both LUCIs house the same complement of broad-, 
medium-, and narrow-band filters, but do not have the same spectroscopic gratings.   
Currently, DL AO spectroscopy is not available (and would only be available on LUCI-2).  
ESM offers an opportunity to improve both the angular and spectral resolution of longslit 
and multi-object slit (MOS) spectroscopy by improving the signal-to-noise (S/N) 
and allowing the use of smaller slitwidths.  For more information regarding the available 
filters, gratings, etc, see Rothberg et al. 2018\cite{2018SPIE10702E..05R} and references
therein.

\section{Characterization of ESM}\label{sec:charESM}
\subsection{Early Testing}\label{subsec:earlyESM}
\indent Using only low order modes to improve angular resolution was a capability already
part of the FLAO system from it's inception (Esposito et al. 2011\cite{2011SPIE.8149E..02E}
and Esposito et al. 2012\cite{2012SPIE.8447E..0UE}).  This mode of operation is benefited 
by the large angular size of the AO Reference Star patrol field, which covers a sizable
fraction (more than 1/3$^{rd}$) of the full LUCI FOV.   The first on-sky test of something
resembling ESM was reported in Esposito et al. 2011\cite{2011SPIE.8149E..02E}. The FLAO 
system was operated in bin mode $\#$4 using only 10 modes with an AO reference 
star of {\it m}$_{\rm R}$ $=$ 13.2 but with a neutral density filter to simulate a star
of {\it m}$_{\rm R}$ $\sim$ 17.2  The observations were conducted using the InfraRed Test
Camera (IRTC - Foppiani et al. 2011\cite{2011ExA....31..115F}).  The measured {\it H}-band 
full-width at half maximum (FWHM) of the star was 0\arcsec.24, while open-loop (SL) 
yielded an {\it H-band} FWHM $=$ 0\arcsec.41 (see Section 4.2 and Figure 6 of 
Esposito et al. 2011\cite{2011SPIE.8149E..02E}).\\
\begin{figure} [H]
\begin{center}
\includegraphics[height=5.2in,width=5.2in]{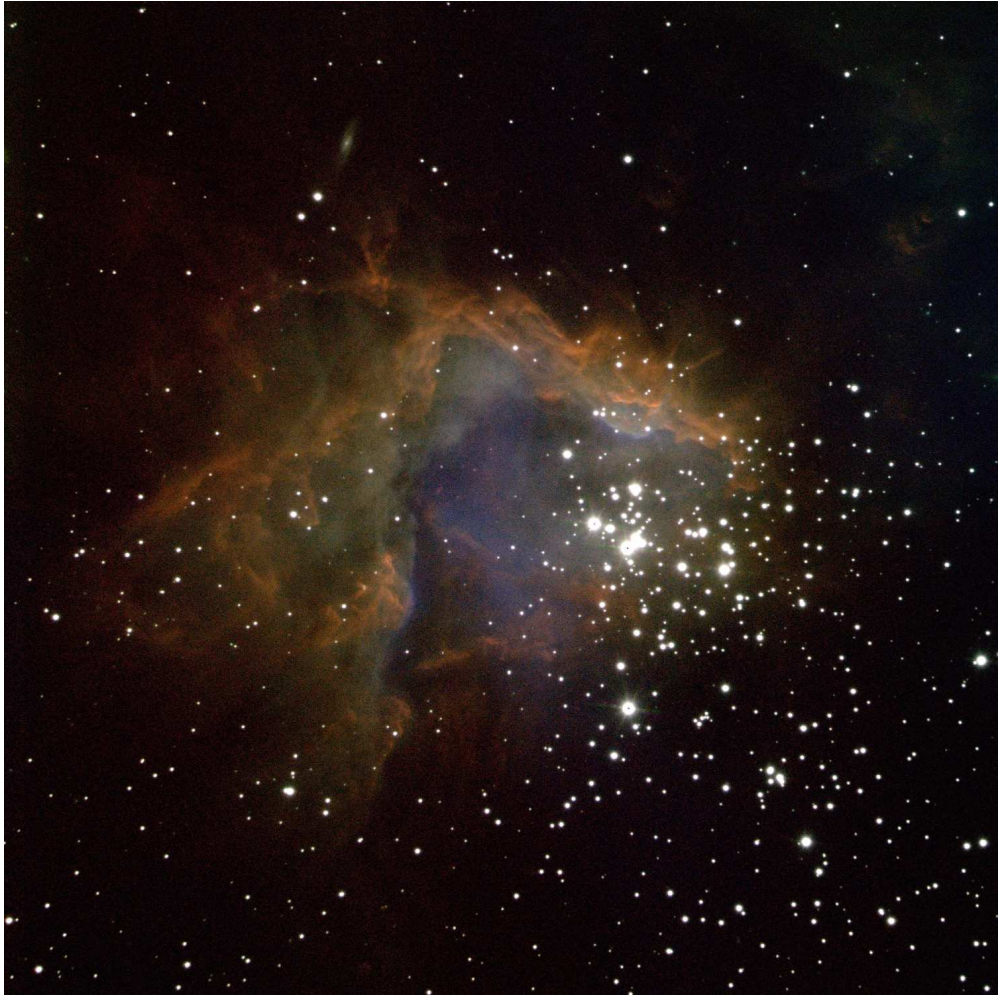}
\end{center}
\caption{False color image of the reflection nebula NGC 1931 created from Brackett 
$\gamma$ (blue - 160 seconds),  {\it K}$_{\rm s}$ (green - 192 seconds), and H$_{\rm 2}$ 
(red - 120 seconds) filters on LUCI-1 using ESM. The natural seeing at {\it K}$_{\rm s}$ 
was $\sim$0\arcsec.52 and $\sim$0\arcsec.25 with ESM. Data were obtained on UT 2004-11-19
and processed by D. Thompson.}
\end{figure} 
\indent Subsequently, ESM was formalized to include 11 modes of corrections in conjunction
with the N3.75 camera on LUCI-1 (before LUCI-2 was commissioned) in order to provide
improved angular resolution across a wider FOV.  Any star can be used as an AO Reference star
as long as it is {\it m}$_{\rm R}$ $\le$ 16, falls within the AO patrol field, and does 
not significantly impact the LUCI detector (i.e. saturate or leave strong persistence).
The next tests of ESM were performed on-sky using LUCI-1 imaging in the fall of 2014.  The 
targets included several bright stars and the nebula NGC 1931, which contains a number of 
young stars with a range of magnitudes fairly well distributed over the LUCI FOV.  
Figure 1 shows a false color composite image comprised of images taken with the 
Brackett $\gamma$, H$_{\rm 2}$, and {\it K}$_{\rm s}$ filters.  Analysis of the 
{\it K}$_{\rm s}$-band data showed a factor of two improvement over the natural seeing.  
Subsequently, ESM imaging was only been attempted a few times during nighttime science 
operations with LUCI-1.

\subsection{Binocular Testing}\label{subsec:binoESM}
\indent Starting in May of 2018, LBTO began a more rigorous characterization of the
capabilities of ESM using imaging data.  This included binocular LUCI-1 and 
LUCI-2 observations in order to:  1) obtain SL and ESM data to compare the 
natural seeing with ESM simultaneously; 2) simultaneous ESM observations to determine 
whether any significant differences exist between LUCI-1 and LUCI-2 which may affect ESM 
image quality; 3) observe a range of AO Reference Stars to determine the impact of 
brightness on the applied corrections; 4) observe stellar fields (globular and open 
clusters) which provide uniform coverage across the LUCI FOV and cover a range of 
magnitudes; and 5) observe large resolved objects to quantify improvements as a function 
of brightness of, and distance from, the  AO Reference Star.  Early results were presented 
in Rothberg et al. 2018\cite{2018SPIE10702E..05R}. Since then additional observations 
were obtained and analyzed.  The characterization project was expanded to 
include {\it J}-, {\it H}, and {\it K}-band data in order to quantify the improvements 
across the full wavelength range of the LUCIs and to include revisiting the same fields 
in order to quantify ESM performance under different natural seeing conditions as well
as repetition of performance under the same conditions.
Presented here are a broader range of multi-wavelength results.  These improvements should 
equally benefit spectroscopic observations as ESM delivers the improved PSF to the focal 
plane unit of the LUCIs in which either spectroscopic masks (longslit or MOS) sit or no 
mask if imaging mode is used.

\subsubsection{Improvements in Image Quality at {\it K}-band}\label{subsubsec:ESMIQ}
\indent {\it K}-band observations of the north-east (NE) periphery of the Milky Way 
globular cluster M92 were obtained over two non-contiguous nights using both LUCI-1 and 
LUCI-2 under seeing conditions 
that ranged from 0\arcsec.70-2\arcsec.06 as measured by the Differential Image Motion 
Monitor (DIMM). These values are corrected to zenith at 0.55$\mu$m.  ESM data were obtained 
using AO Reference Stars with {\it m}$_{\rm R}$ $=$ 9.43 on both nights and 
{\it m}$_{\rm R}$ $=$ 15.17 on the second night.  
ESM was turned on for one set of data, then the observations were repeated under SL 
conditions (ESM off). The data were reduced using {\tt IRAF}. The reduction process 
includes linearization corrections, bad pixel masking, flat-fielding, background 
subtraction.  The {\tt IRAF} tasks {\tt geomap} and {\tt geotran} were used to rectify 
each exposure by applying sub-pixel shifts, rotation, and accounting for any distortions 
from the LUCI optics.  The {\tt IRAF} task {\tt daofind} was then used to identify stars 
in the crowded field for each final mosaic image.  Non-astronomical objects and stars 
near the edges of the FOV were removed using an automated {\tt IDL} routine.  The 
{\tt IRAF} task {\tt radprof} was then used to measure a Gaussian FWHM for every star.  
For each star the radial distance from the AO Reference Star was computed. Figure 2 shows 
a plot of the FWHM in angular units of arcseconds plotted against the radial distance from 
the AO Reference Stars for the two epochs.  The left panel shows data SL data and ESM data
obtained under poor seeing with LUCI-1.  The right panel shows a similar comparison, 
but taken under excellent natural seeing conditions and using two different AO Reference 
stars.  The left panel shows that under poor seeing conditions ESM can ``recover'' 
a night and improve the image quality delivered to the LUCI focal plane by almost a factor
of three.  Data obtained with LUCI-2 show the same improvement.  Additional data taken
of other fields (the Milky Way open cluster M52, the Milky Way globular cluster Palomar 2 
- see section~\ref{subsec:caseESM}), as well as additional data obtained of M92 show
this improvement is repeatable.  The right panel in Figure 2 demonstrates that under
optimal conditions ESM can achieve a resolution of 0\arcsec.23 or better ($\sim$ 2 pixels
with the LUCI N3.75 camera).  The right panel also shows that the corrections achieved
with ESM have a slight dependence on the brightness and distance from the AO Reference 
Star.\\
\begin{figure} [h]
\begin{center}
\includegraphics[height=2.78in,width=6.75in]{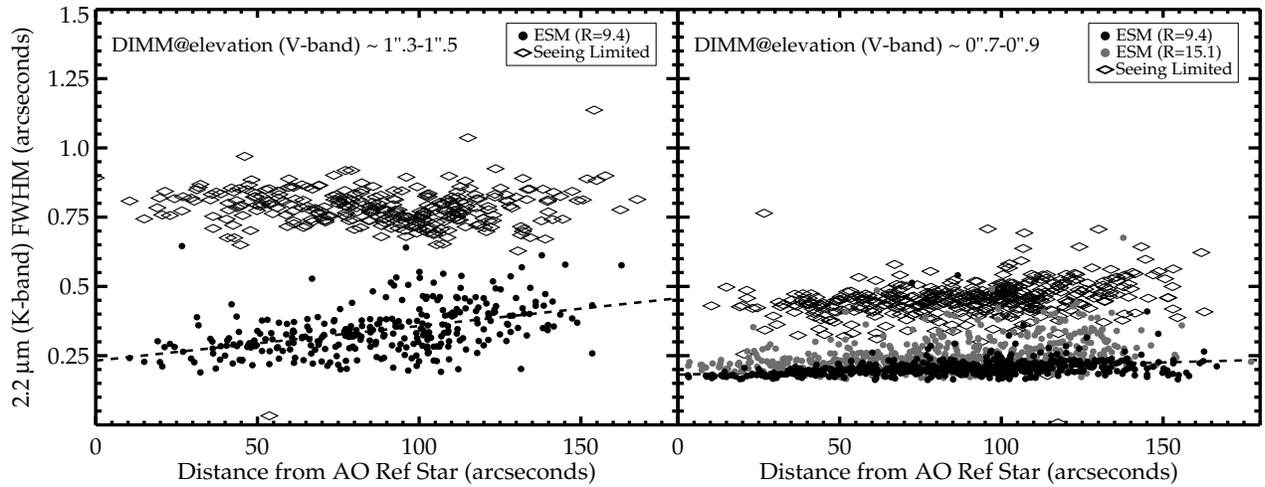}
\end{center}
\caption{Comparison of SL and ESM values of the FWHM of stars in the NE periphery of
the globular cluster M92 as a function of distance from the AO Reference Star (for
SL observations this is only to make a direct comparison - the AO Reference Star has
no impact on the image quality).  The left panel is for data taken under poor natural
seeing conditions and shows a 3$\times$ improvement of the PSF.  The right panel
shows the same field taken under excellent natural seeing conditions and
adds a second, fainter, AO Reference star for comparison.  While the improvement is
only a factor of two over SL data, ESM produces a nearly uniform improved PSF of 
$\theta$ $\sim$ 0\arcsec.23.  The fainter AO Reference star does produce marginally poorer
image quality ($\theta$ $\sim$ 0\arcsec.25).  The dotted line is a linear least-squares
fit to the data to assess the quality of the improvement.  The total integration time
for each SL and ESM final mosaic image on the left is 564.75 seconds, and 376.5 seconds on the 
right.}
\end{figure} 
\begin{figure} [H]
\begin{center}
\includegraphics[height=3.04in,width=6.75in]{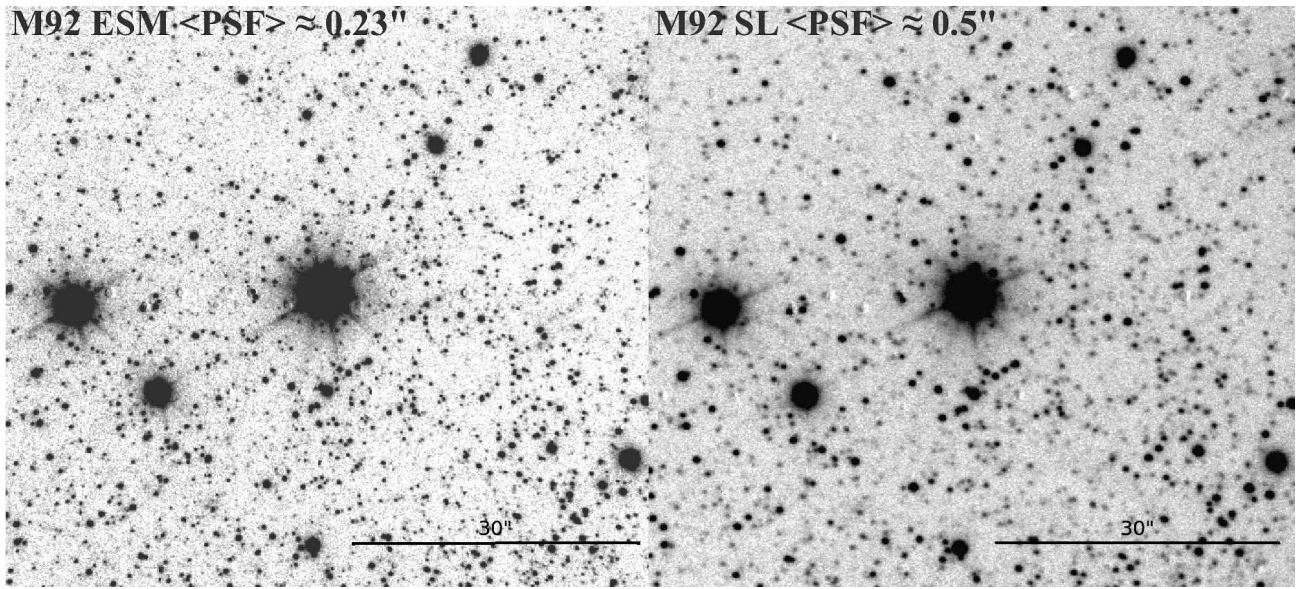}
\end{center}
\caption{Comparison of the central 1$\arcmin$ around the {\it m}$_{\rm R}$ $=$ 9.43 AO 
Reference Star for ESM ({\it left}) and SL ({\it right}) {\it K}-band data of the NE 
periphery of the Milky Way globular cluster M92.  The figure provides a qualitative 
comparison of the improvement of the PSF and depth achieved in optimal conditions by 
ESM compared to SL data obtained under excellent natural seeing conditions.}
\end{figure} 
\indent The comparison between SL and ESM observations shown in Figure 3 
demonstrates the clear improvement in sensitivity of ESM over 
SL observations. The total integration time for each set of {\it K}-band observations 
was 376.5 seconds.  Both observations used the same dither pattern, and were 
processed in the same way using {\tt IRAF}.\\

\subsubsection{{\it J}-band Image Quality}\label{subsubsec:resolvedESM}
\indent  As part of the characterization of ESM using large resolved objects and 
observations at $\lambda$ $<$ 2$\mu$m observations of the advanced (single 
nucleus) merger remnant NGC 3921 were obtained under poor seeing conditions.  NGC 3921 was 
selected from a larger sample of 51 advanced merger remnants 
(Rothberg \& Joseph 2004\cite{2004AJ....128.2098R}). The angular sizes of these objects
are all at least 1$\arcmin$ in diameter and have published or archival multi-wavelength
ground- and space-based imaging datasets available 
(i.e. Rothberg \& Fischer 2010\cite{2010ApJ...712..318R}) allowing for comparisons with
ESM performance.  NGC 3921 was selected based on visibility and a nearby AO Reference 
star.  \\
\begin{figure} [H]
\begin{center}
\includegraphics[height=5.75in,width=5.75in]{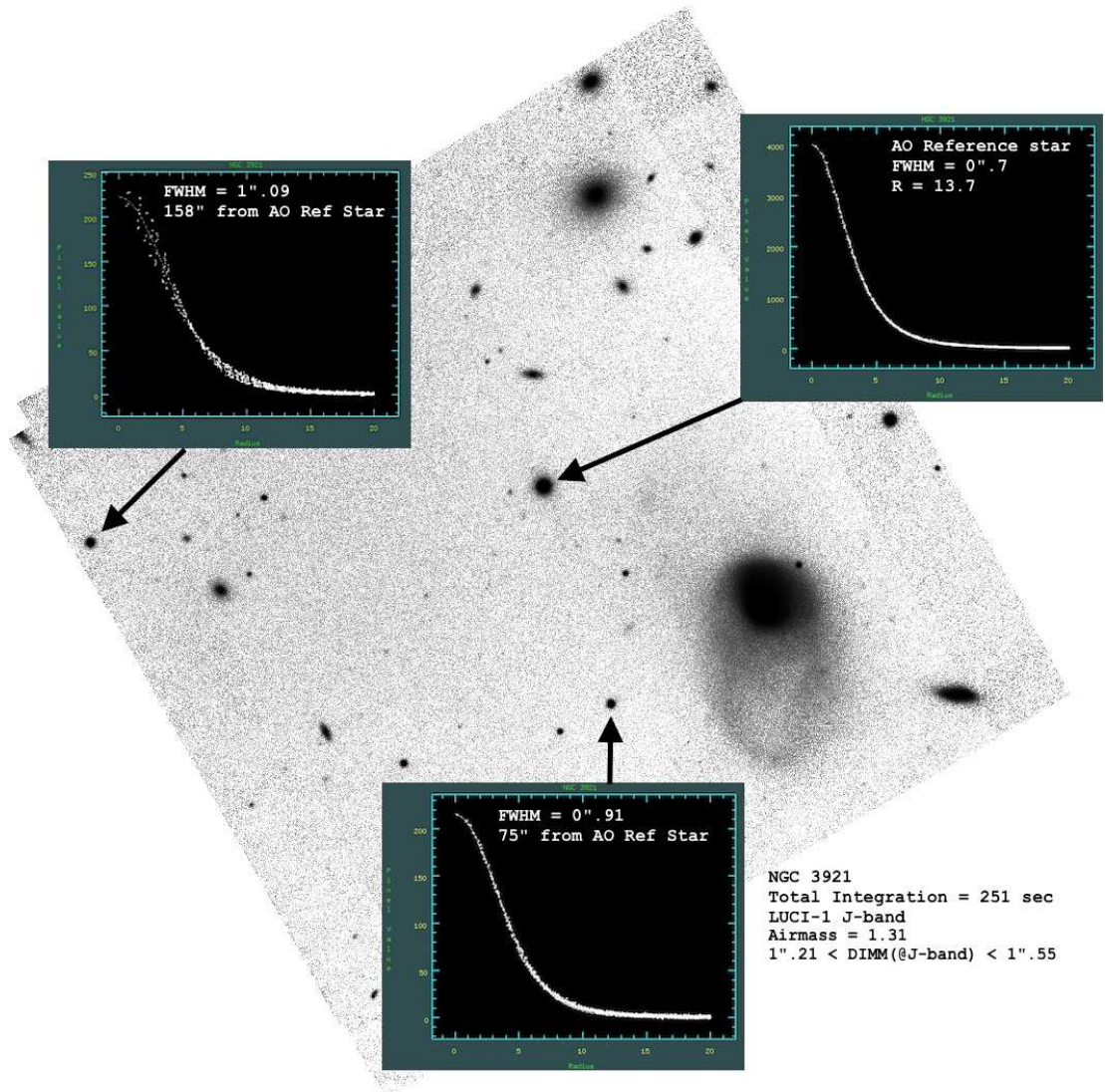}
\end{center}
\caption{{\it J}-band (1.25$\mu$m) final mosaic image of the advanced merger remnant 
NGC 3921 obtained with LUCI-1 using ESM with a {\it m}$_{\rm R}$ $=$ 13.7 AO Reference 
star.  Shown inset are 1D light profiles of various stars in the field, including the AO 
Reference Star.  The DIMM seeing, corrected to {\it J}-band and the elevation of the target 
is shown in the figure.  In poor seeing ESM can deliver a factor 70-200$\%$ improvement 
to the LUCI focal plane at {\it J}-band. }
\end{figure} 
\indent Preliminary results clearly demonstrate that the image quality
at {\it J}-band can significantly benefit from the use of ESM, even in poor conditions.
Figure 4 shows a mosaiced field of NGC 3921 created from four exposures of 62.75 seconds, 
processed and combined using {\tt IRAF} as described earlier.  The data were taken 
under natural seeing (corrected to {\it J}-band and the elevation of the target) of 
1\arcsec.21-1\arcsec.55.  The 1D light profiles of stars show
that the corrections improved the image quality to $\theta$ $=$ 0\arcsec.7 for
the AO Reference Star and to $\theta$ $\sim$ 0\arcsec.9 near the position of 
NGC 3921.  This type of significant improvement demonstrates the ability to obtain 
sub-arcsecond quality science data in poor conditions over the full wavelength range
of the LUCIs. \\

\section{Scientific Case Study for ESM - Palomar 2}\label{subsec:caseESM}
\indent As part of first science demonstration case to test the improvements ESM delivers
to the LUCI focal plane, the Milky Way globular cluster Palomar 2 was selected for
observations at {\it J}-, {\it H}, and {\it K}-band. Palomar 2 lies near the anti-center 
of the Milky Way (relative to Earth) and behind significant patches of interstellar dust. 
First discovered in 1955 by Abell\cite{1955PASP...67..258A}, it remains one of the least 
studied GCs. The first distance estimate and stellar census were obtained by 
Harris (1980)\cite{1980PASP...92...43H}.  To date, only optical observations of Palomar 2
have been published.  LUCI near-IR observations of Palomar 2 were obtained over two 
non-contiguous nights.  ESM observations could only be obtained with 
LUCI-2 (DX) due to the ongoing commissioning of the SOUL upgrade on SX. SL observations 
were obtained simultaneously with LUCI-1 on one night allowing for a direct comparison 
between the natural seeing and the image quality from ESM.  
Only LUCI-2/ESM observations were obtained on the first night.\\
\begin{figure} [H]
\begin{center}
\includegraphics[height=3.73in,width=6.6in]{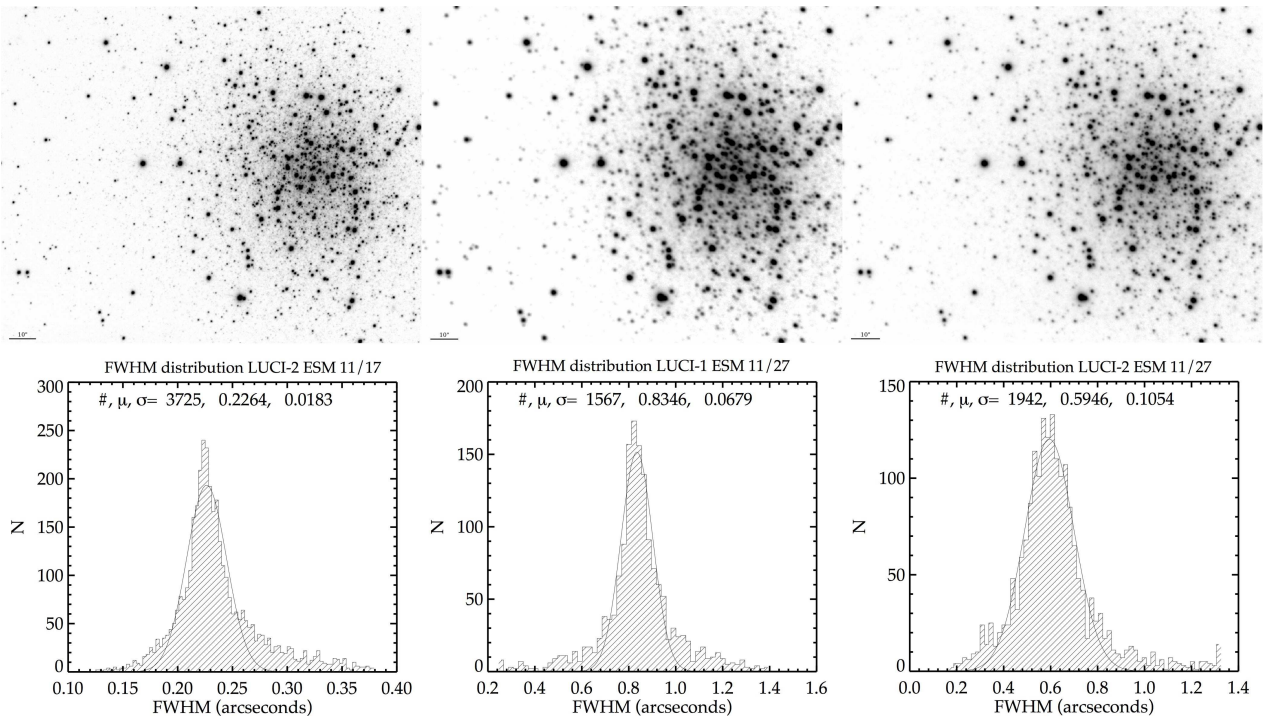}
\end{center}
\caption{{\it Top Row}: Comparison of K-band (2.2$\mu$m) images of Palomar 2 obtained with 
{\it left} - LUCI-2/ESM - observations with an average {\it K}-band FWHM $\sim$ 0\arcsec.23 
and an average {\it V}-band DIMM seeing of 0\arcsec.86 or 0.\arcsec65 (corrected to 
{\it K}-band and to the elevation of the observations); {\it center} - LUCI-1/SL with an 
average {\it K}-band FWHM of 0\arcsec.83 and an average {\it V}-band DIMM seeing of 
1\arcsec.3 (corrected to the elevation of the observations); {\it right} - LUCI-2/ESM
observations taken simultaneously with the LUCI-1/SL ({\it center}). The average 
{\it K}-band FWHM of 0\arcsec.59.  {\it Bottom Row}: The distributions of the 
FWHM measured for all of the objects in each {\it K}-band image shown above it. Each 
histogram shows the final number of stars (after sorting to remove spurious detections),
the average FWHM in arcseconds ($\mu$), and the 1$\sigma$ errors in the distribution.}
\end{figure} 

\indent The first part of the analysis measured the image quality of the data.  The
science frames were processed, aligned, and a final mosaic was created using the {\tt IRAF} 
procedures described earlier. For the data taken on the 2nd night (UT 2018-11-27), 
the SL and ESM data were obtained simultaneously allowing for a direct measurement of 
the real-time improvement of the image quality.  This removes any variability or 
uncertainty from the atmospheric turbulence since the same Cn2 (optical turbulence) 
profile is applicable to both the SX and DX sides of the telescope.  A distribution of
the {\it K}-band FWHM of the stars in Palomar 2 are shown in Figure 5 for each set of
data (bottom row) along with the image of Palomar 2 associated with it (top row).  The
AO Reference Star used for the observations was listed as {\it m}$_{\rm R}$ $=$ 12.86 
and was 141$\arcsec$ away from the core of Palomar 2.  However, upon acquisition, 
it was discovered the ``star'' was actually a double-star (separated by 1\arcsec.5), 
but this did not affect closing or maintaining the AO loop each night.  The results were 
consistent with previous observations, in that data obtained under excellent natural 
seeing conditions yielded a FWHM $\sim$ 0\arcsec.23$\pm$0.01, even with such a large 
angular separation between the (double) AO Reference Star and the core of Palomar 2.\\
\indent Under average seeing conditions during UT 2018-11-27, ESM produced a significant
improvement in the PSF.  The FWHM improved by $\sim$ 1.4$\times$, from
$\theta$ $\sim$ 0\arcsec.83$\pm$0.06 measured with LUCI-1 to 
$\theta$ $\sim$ 0\arcsec.59$\pm$0.1 measured with LUCI-2.  Furthermore, the LUCI-2 stellar
PSFs were rounder than the SL PSFs of LUCI-1 (see Figure 5).\\ 
\begin{figure} [H]
\begin{center}
\includegraphics[height=4.5in,width=4.5in]{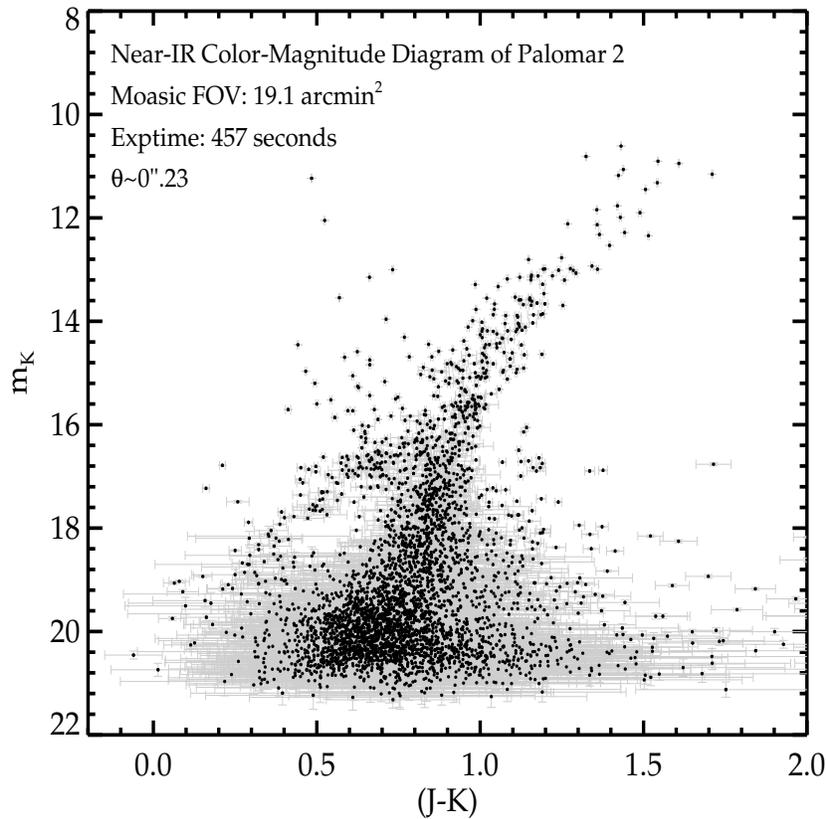}
\end{center}
\caption{Near-IR Color-Magnitude Diagram (CMD) for Palomar 2 based on data obtained with
ESM on UT 2018-11-17. In only 457 seconds, the data are able to achieve a remarkable
depth probing to the level of the main sequence turnoff.  Each point plotted is shown
with error bars. }
\end{figure} 
\indent In addition to assessing the image quality, the data obtained with LUCI-2/ESM on 
the first of the two nights were used to construct a near-IR color-magnitude diagram (CMD) 
of Palomar 2.   The {\it J}-,{\it H}-, and {\it K}-band images were aligned with each
other using the {\tt IRAF} tasks {\tt geomap} and {\tt geotran}.  The {\tt IRAF} task 
{\tt daophot} task was used to find sources in the crowded field and the task 
{\tt radprof} was used to measure the FWHM, magnitudes and magnitude errors.  The data 
were first sorted to remove objects with null or INDEF values and magnitude errors 
$\ge$ 0.3 mags.  The near-IR CMD reaches $\sim$ 1 magnitude deeper in the same integration 
time (457 seconds in total for each filter) as data obtained from the SL and ESM 
observations obtained under average conditions.  The near-IR CMD is shown in Figure 6 using
{\it J}-, and {\it K}-band data.  It clearly shows features such as the main sequence 
turnoff, asymptotic giant branch stars, and a clump of red giant branch (RGB) stars. 
The tip of the RGB can be used to infer a distance.  Further analysis of this data
will be performed and the results presented in a future publication.  However,
the preliminary results clearly show the power of ESM under excellent seeing conditions
in combination with the 4$\arcmin$ $\times$ 4$\arcmin$ LUCI FOV.\\

\section{Conclusions \& Future Work}\label{subsec:ConFut}
\indent The results, to date, indicate that ESM provides a significant improvement in 
image quality regardless of the natural seeing conditions or the the wavelength 
(1-2.4$\mu$m) observed.  While some improvement is expected, the most unexpected result 
from the analysis so far is the extent of the improvement observed in extremely poor 
conditions (i.e. $\theta$ $>$ 1\arcsec.5).  The philosophy of the current characterization 
of ESM is to revisit the same fields multiple times regardless of the atmospheric 
conditions.  This removes systematics (analyzing the same targets in the same way, 
thus comparing daleks to daleks not daleks to apples) and provides an opportunity to either
sample different atmospheric conditions or sample the same atmospheric conditions to 
verify repeatability of results.   In the best natural seeing conditions, the {\it K}-band 
image quality frequently approaches the Nyquist limit of the N3.75 camera.   
The conclusion thus far is that as long as long as there is a viable AO Reference Star 
in the patrol field and the FLAO (or in the future, SOUL) system can 
close and maintain the AO loop, {\it all} LUCI observations should use, and would 
benefit from, ESM.\\
\indent The characterization of ESM is still an ongoing process.  Although multi-epoch 
{\it J}-, {\it H}-, and {\it K}-band data have been collected and reduced for M92 and M52, 
additional analysis remains to be completed (this includes the corrections achieved with
an AO Reference Star at the magnitude limit).  In addition to NGC 3921, multi-epoch
observations of Hickson Compact Group 31 (HCG 31) have been obtained.  Like NGC 3921, 
the goal is to determine the improvements ESM can provide to resolved objects as
a function of the brightness of, and distance from an AO Reference Star.  However, HCG 31 
contains at least 5 discrete, resolved, galaxies at varying distances from the 
{\it m}$_{\rm R}$ $=$ 10.9 AO Reference Star within the LUCI FOV.  These analyses will
be presented in a future publication.\\
\indent With the installation and commissioning of SOUL on SX and the upcoming installation 
of SOUL on DX, ESM characterization will assess the improvements delivered by these 
upgrades.  Of particular importance is the ability to use fainter AO Reference Stars to 
provide  corrections.  SOUL should allow for the use of an AO Reference Star as faint as 
{\it m}$_{\rm R}$ $\sim$ 17.5-18.  This opens up a much larger part of the sky for ESM
observations and improves the likelihood of finding a suitable AO Reference Star in any
given LUCI science field.  This will need to be quantified.  As shown in Figure 2, there 
is a weak dependence on the corrections achieved in the image quality as a function of the 
brightness of the AO Reference Star.  Further ESM characterization will be needed to 
determine the impact of this with the new AO Reference Star limits.


\acknowledgments 
The authors would like to thank R.T. Gatto for useful discussions and providing much 
needed support in the planning and analysis of the work presented.


\end{document}